**Title:** Evidence for strong modality-dependence of chronotype assessment from real world calendar app data

**Authors**: Sourabh Gapate[1], Royan Kamyar[2], *Benjamin L. Smarr[3,4]
1. Sourabh's program (Master of Science in Business Analytics - Rady School of Management, University of California San Diego, La Jolla, CA, USA)
2. Owaves Inc
3. Shu Chien - Gene Lay Department of Bioengineering, University of California San Diego, La Jolla, CA, USA
4. Halicioglu Data Science Institute, University of California San Diego, La Jolla, CA, USA
*Corresponding author; bsmarr@ucsd.edu

## Abstract

Chronotypes allow for the comparison of one individual's daily rhythms to that of others, as well as to their environment. Mismatch between an individual's chronotype and the timing constraints of their social environment create social jet lag, which is correlated to mental and physical health risks. Tracking chronotypes by assessing relative daily phase of pre-existing data from diverse sources is one way researchers have attempted to capture the extent and impacts of social jet lag in the real world. The concept of chronotype implicitly supposes that a single phase applies to an individual, whereas the circadian rhythms of different internal systems entrain to or have their outputs masked by different environmental inputs, at least somewhat independently. Therefore, it is worth testing the extent to which the timing of data associated with the output of one such internal system predicts the timing of the others within the same individual in a modern context. If it is true that the modern environment interferes with internal synchrony and/or generates different masking for different internal systems' outputs, then real world data reflecting these outputs through different behaviors ought to reveal relatively low correlations, reflecting environmental interference. At the other extreme, if there is no behavior or tissue specific interference, then different internal system outputs should all be equally predicted from data modalities capturing their different outputs.

Here we explore multimodal behavioral rhythm data from the Owaves calendaring app, focusing on behavioral outputs logged as: Sleep, Exercise, Eat, Work, Love, Play, Relax, Misc. We find that individuals show daily rhythms within each behavior type from which chronotypes can be assigned, but that chronotypes derived from different behaviors (or combinations of behaviors) lead to nearly independent sorting of individuals by phase. This suggests that if real world data are used to assign an individual's chronotype, then that chronotype may be specific to the internal system the output of which is related to each specific data modality assessed. Our findings suggest that researchers cannot confidently assume to account for outputs from other internal systems in real world settings without additional observations or controls.

## Introduction

Historically people have been classified into chronotypes to identify the phase of their circadian rhythms relative to others. Chronotype assessments have been used to categorize individuals using a range of data types, from digital school logs[1], [2] to social media activity[3], [4] to one time retrospective questionnaires[5], [6]. These in turn are abstractions from clinical assessment tools such as Dim Light Melatonin Onset (DLMO) [7], [8]. Chronotype is often discussed as a unitary trait belonging to an individual, with stereotyped changes occurring across life, as well as with demographic factors[9], [10]. However, many studies have also sought to compare the estimates of various approaches, which tend to correlate, but not with extreme precision (e.g. [11], [12]; the full review or even meta-analysis of that literature would be quite interesting, but is beyond the intended scope of this manuscript).

Circadian regulation is not relegated to one body part [13], [14], but seems to have evolved within individual cells from which multicellular organization arose later [15], [16]. It seems more appropriate to think of individuals as coupled oscillator networks[17], [18] composed of cells and tissues each with their own intrinsic and entrainable rhythms. The resulting hypothesis that humans are coupled oscillator systems makes two important predictions for the work presented here: 1) individuals do not have "a phase" but represent many phases that, under healthy conditions, share stable alignments; 2) the outputs of different internal systems may react differently to different environmental and social pressures. This framework suggests that different body system-related outputs might entrain to different zeitgebers (e.g. [19], [20]) or be masked differently by specific environmental cues (e.g. [1]), leading to less reliable phase alignment within an individual's various outputs when in modern social contexts. From the point of view of using diverse data sources to infer chronotype for an individual, this framework suggests the need for testing as to how consolidated / stably aligned different output phases really are within an individual, and what that means in terms of whether an individual should be thought of as having "a chronotype" as opposed to "a chronotype for sleep," "for exercise," "for eating," etc.

In this study we examine the hypothesis that data generated in the real world by the timing of different human outputs stably align within an individual, consistently reflecting that individual's chronotype. We leverage daily rhythm data from Owaves, a day planner app, capturing self-reported timing records of many different outputs (Sleep, Exercise, Eat, Work, Love, Play, Relax, and Miscellaneous activity labels) from 1M+ activities across 229 individuals longitudinally. We employ multiple clustering and classification approaches to test the null hypothesis that chronotype classification is not meaningfully affected by the choice of activity used to generate the classification.

## Results

Data filtering (see Methods) resulting in a pool of **229** individuals with sufficient data for inclusion in these analyses. The composition of this populations was **13** males, **96** females, **13** others, **107** unknowns; mean age **26.5 +/- 9.5 years** standard deviation. We did not have associated ethnicity data for these individuals, but it is presumed to be majority white gathered from the United States.

Here we compared eight different kinds of output from the same people who logged the time of different outputs in the Owaves calendar app. We sorted all individuals within each output by the median time of each output's daily midpoint (e.g., an individual's median midsleep phase) to create chronotype proxies for each individual within each output. For example, **Fig. 1A** shows three chronotypic sortings across all individuals (we show only a few examples of the 8x8 possible comparisons for the sake of brevity in the illustration, but report all pairwise comparisons in the tables). All within-output sortings showed a similar capacity for enabling assignment of chronotypes to individuals, visible as a diagonal line of individual phases progressing from the top left to the bottom right in **Fig. 1A**). When visually assessing the ranking across outputs – the rank order of one output visualized using the individuals' phases from any other output – that capacity for assigning chronotypes disappears. **Fig. 1B** illustrates that the rank order across different outputs appears random, suggesting that the timing of one activity (e.g., sleep) does not predict the timing of another (e.g., exercise) within the same individual. This lack of correlation across output-related data modalities indicates that different behavioral chronotypes are largely independent. The loss of a clear line progressing from one row to the next reflects a lack of relationship between rows of individual's phase rank from one output (Y axis) versus the timing of events in a different output (X axis).

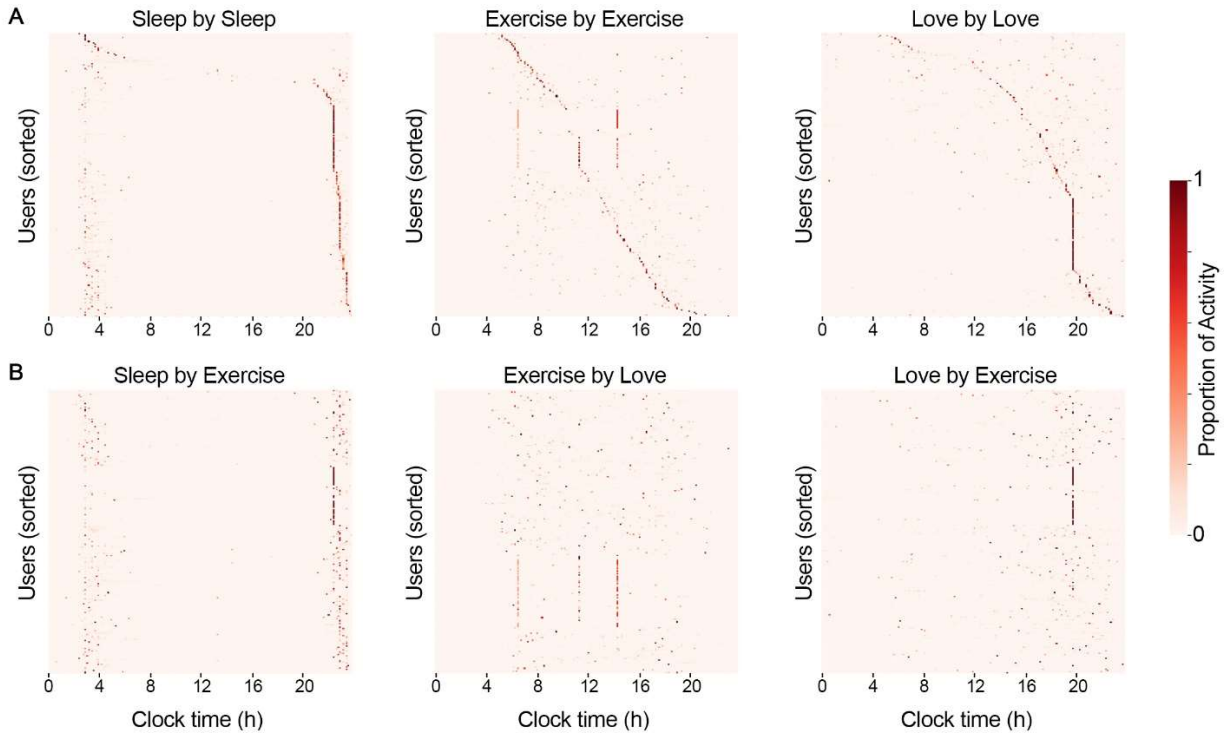

**Figure 1. Comparison of rankings as chronotypes by one modality are not preserved into other modalities.** A: comparison of sleep timing (the most common chronotype proxy) to exercise timing (a common proxy) and love (a novel proxy). Within each there are a range of timings consistent with the concept of chronotype (A). When sorted by each other (B) the range is lost, suggesting that the information from one output has a relatively random predictive value of the same people's timing in other outputs; colors are arbitrary within each graph and not suggest relationships across panels.

We then sought to quantify the similarity of chronotypes from each output to each other output; to test the hypothesis that each individual should have a similar chronotype regardless of the specific output used as a proxy for their internal circadian rhythms. We directly measured resultant chronotype similarities by Adjusted Rand Index (ARI) to test for related vs. random sorting across outputs, using two forms of clustering to ensure the results were not dependent on our choice of clustering algorithm. In both hierarchical clustering on raw data (**Fig. 2A**) and K-means on principal component projections of the data (**Fig. 2B**), ARIs were low for all cross-output comparisons (**Fig. 2C,D**). By contrast, when we introduced random but systematically increasing disorder into a given hierarchy, the ARI started high and drifted lower with increasing disorder, as expected. This confirms that the ARI metric is operating as expected, but that clusters generated within each output show little relationship to each other across outputs (**Fig. 2E, inset in C**). The randomized ARI after 100 trials (100 randomizations from the original 229 individuals) varied between 0.25 and 0.45. The mean and median ARI for hierarchical clustering were 0.07 and 0.03; for K-means these were 0.09 and 0.07. This reflects the ARI of lists by assessing the matching of exact ordering user by user. This low rate of agreement remained the trend when using composite modalities of either the 4 most commonly used modalities (sleep, eat, exercise, work) or of all modalities.

ARI compares the relative positions of the ends of each branch (individual users). We therefore additionally compared the proportion of users within each cluster that remained in the same cluster when compared across modalities (clusters are grouped by color in **Fig 2A,B** – colors do not correspond across plots, as color is assigned per cluster, the order and composition of which changes with each different analysis). This lowered the propensity of the above ARI metric to overvalue small changes in rank that might not reflect change in real-world chronotype categorization (e.g. owls rank 1 and 3 out of 100 could switch places and lower ARI in our first analysis, but that small of a change would not likely change both of their classifications from owl to something else). We calculated the proportion of IDs preserved across activity type pairs by aligning clusters to maximize shared individuals. Then for each cluster in one activity type, we aligned it with the cluster in each other activity type that shared the maximum user overlap, and calculated the proportion of users preserved (**Fig 2F**). We used randomized cluster assignments to establish a baseline comparison **Fig 2G**). Using these broader groupings, we found generally higher agreements with a mean 48% and a median 45% preserved per cluster across activity types. By comparison, the randomized baseline showed mean 45% and median 43% preserved at chance, which is slightly lower than the percent we found in the actual cluster comparisons, but not substantially different.

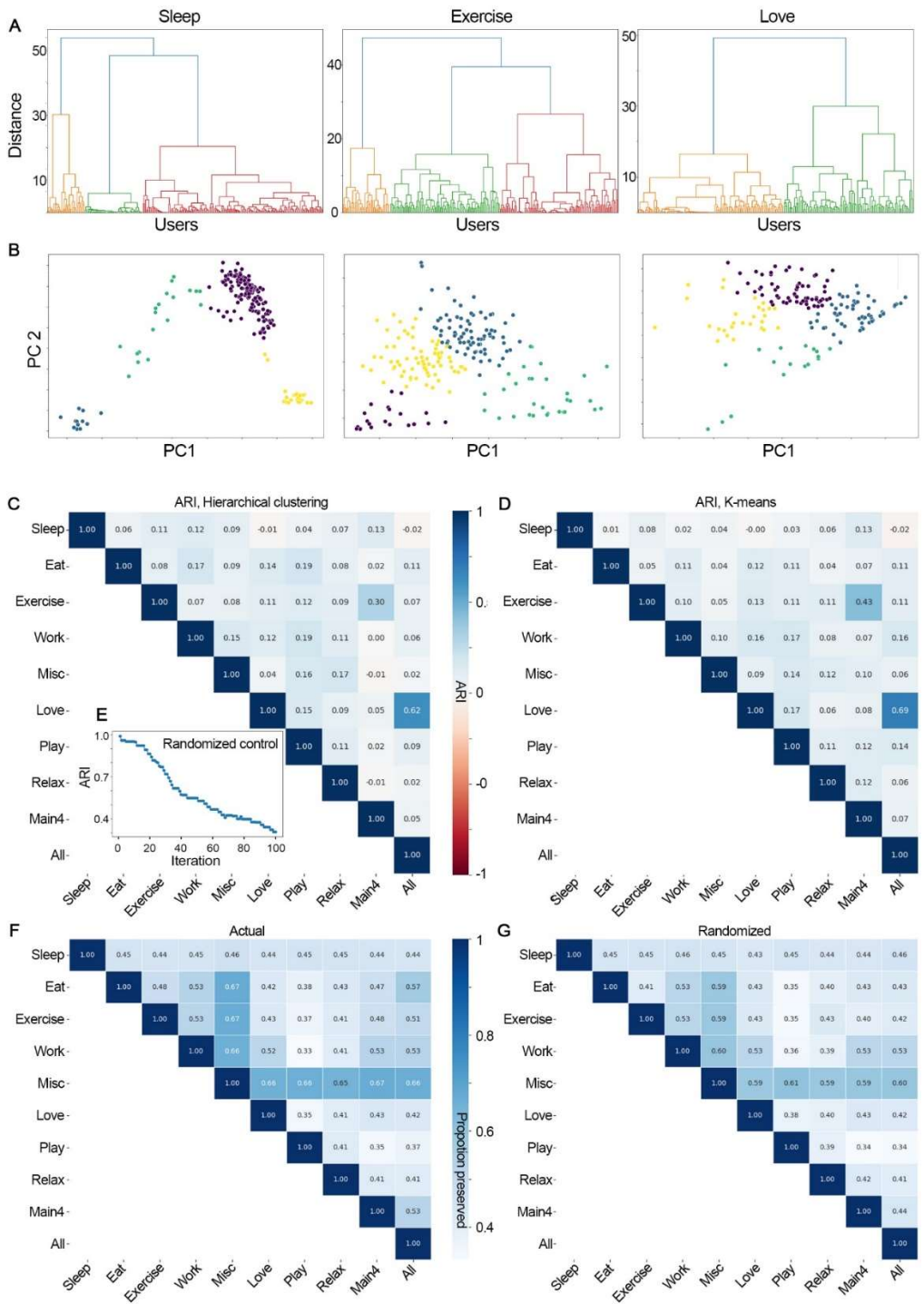

**Figure 2. Numerical comparisons confirm a lack of chronotype transfer from one modality to another.** Both Hierarchical (A) and K-means (B) clustering (here showing the same example modalities as in Fig. 1) confirm a substantial differences in sorting by different outputs. Across all eight modalities in this data set, ARIs confirm the relatively random relationship between chronotypes assigned by any modality to that assigned by any other (C, D). As a positive control (E), ARI values decrease from 1 toward 0 proportionate to the number of pairwise random swapping of user ranks from a given hierarchy. Proportion of preserved within a shared cluster (F: hierarchical) was still near that of chance (G). Note, the higher proportion of alignment with Misc in both Actual and Randomized suggests a lack of independence of Misc from other modalities, rather than a meaningful use case for chronotype assignment.

## Discussion

In this manuscript we found that all 8 different behavioral data modalities assessed could be sorted to allow for individual assignment of chronotype. When comparing the rankings by phase across modalities, in almost all cases the relationships were nearly random. Assessing rank only as likelihood of individuals still being located within the same large clusters (simulating chronotypic "types" like lark or owl) across data modalities improved our findings from about 5% agreement to about 50% agreement, which was similar to but slightly higher than chance. This low level of agreement across data modalities within individuals was consistent by multiple analytic approaches.

This is consistent with 1) either desynchrony of phase across physiological system's outputs and/or 2) differential masking of different outputs within an individual living in the real world. It appears that while each data modality reveals a range of daily timing across people, the relative position or phase of this timing is not stable across data modalities associated with different outputs within individuals. Concretely, we found that in this dataset, saying someone is the earliest sleeper carried little to no information about whether they were the earliest eater or exerciser or anything else. Examining a large set of data from many different real world activity logs, we find that the choice of activity upon which to base assessments of chronotype is highly influential on the outcome of the classification. That is, depending on the activity used as a proxy for circadian rhythms classification into chronotypes, the same individual will have a relative phase that in most cases has a close to random relationship to the same individual's phase assessed by a different data modality.

We cannot say from these analyses whether any one of these eight modalities assessed here is closer to an internal circadian rhythm than another; we do not know what is the output of an endogenous oscillator, and what is just a behavioral output, nor are we able to confirm the level at which planned outputs reported in the app correspond to actual actions taken. As a result, we have attempted to couch our findings as being consistent with either internal desynchrony or differential masking or some combination. We furthermore do not know how common such a low level of agreement would be in different data types or different data sets. A key limitation that must be emphasized is our reliance on self-reported data from Owaves, which likely results in biases, such as over-reporting or under-reporting certain activities. Future studies could incorporate objective measures such as wearables to validate self-reported timing. Additionally, our cohort is not broadly representative, but majority women in their 20s, reflecting the user base of this particular multimodal data source. We are strong proponents of diverse inclusion in data collection, and Sex as a Biological Variable analysis, and so it is frustrating to not be able to more deeply compare effects of different groups on our findings this far; though this data source does provide substantial volumes of longitudinal and multimodal data, allowing for the comparisons we carry out, it does not provide the demographic breadth of coverage that would let us assess the generalizability of our results. Stability of cross-activity phase relationships is likely to change with age and life stage. Behavioral choices are also under social constraint, and so are influenced by socioeconomics, diet, and all manner of other conditions and choices, all of which likely also changes by population, age, and other demographic factors. Future work assessing the stability or relatedness of different longitudinal "digital wakes" left by individuals from diverse circumstances would help to clarify the causal roles of different influences on the rate of activity alignments within individuals. More invasive studies may be needed to disentangle which data modalities are more driven by physiological oscillator outputs, and which more driven by masking.

Even with these limitations, our findings of low across-modality agreement are not entirely surprising. Human activities are not generated by internal circadian rhythms in isolation, but are instead heavily influenced by environment and society. Since those external influences may have differing impacts on different body systems (e.g. food may influence the liver more[21], caffeine or anxiety may influence sleep timing more[22], etc), we might expect to see more plasticity in output timing in humans living in a post-industrial context. Nevertheless we were surprised by the extent of the dissimilarity between activities. Our findings suggest that care is warranted when interpreting phases estimated from different data sources, and that more studies should be carried out to understand the connections between outputs (here, activities) and internal circadian mechanisms. If these outputs are good proxies for internal rhythms of related physiological systems (and we have no measurement of that within this analysis), then internal desynchrony maybe more common and of higher amplitude than is commonly assumed; if they are not good proxies, then interpreting findings from real world studies using digital outputs as proxies for circadian rhythms – including our own work[1] – may need to be reinterpreted.  What our analyses prove, if anything, is open to debate on many fronts, and we try to temper our claims accordingly. What these results do cause us to say is that we can reject the hypothesis that chronotype is definitely a stable construct across data modalities. How stable, for whom, and by which measures is important to work out as more work in circadian rhythms builds on emerging data sources[23].

In summary, our data suggest caution and consideration. Circadian biologists studying real world human data should be careful to assess and report from which output chronotype is being assessed; what presumed physiological systems are implicated, and which data modalities are used as proxies. Large, longitudinal, and multimodal data like those generated by Owaves are rare at present but likely to become more common. Here such a data set revealed that the construct of chronotype may be less stable or generalizable than is often assumed. The limitation of inferring endogenous functions from proxy data must be reiterated. At the same time, the potential of different data modalities serving as proxies from multiple different internal systems is tantalizing. Such data may support increased precision when using timing changes to infer health changes, as in longitudinal monitoring for changes in mental health or chronotherapeutics.

# Methods

## 1. Data Collection

The entire analysis was carried out using Python. We received a dataset comprising over a million activity records from the Owaves team, specifically from users who had consented to participate in the research. The total dataset consisted of 1,037,191 rows, detailing eight different types of user activities: sleep, eat, exercise, work, love, play, relax, and miscellaneous. Each row contained information on the start and end times of the activity, its title, type, location, and notes. For security purposes, the data was de-identified before we received it.

## 2. Data Filtering Pipeline

The dataset was filtered to ensure a sufficient variety of daily activities, allowing us to analyze individual circadian rhythms and the impact of different activities on each other. Days were included in the analysis if they met the following criteria:

- Days should be filled at least 75% (>= 18 hours).
- Days should have at least one activity for each of the following types: Sleep, Eat, Exercise, Work.
- Users should have at least 30 days of data to ensure variability in day plans while maintaining a decent number of users.

Post-filtering, the dataset was reduced to 433,732 rows from 229 individuals, accounting for 56.6% of the original, non-duplicated database.

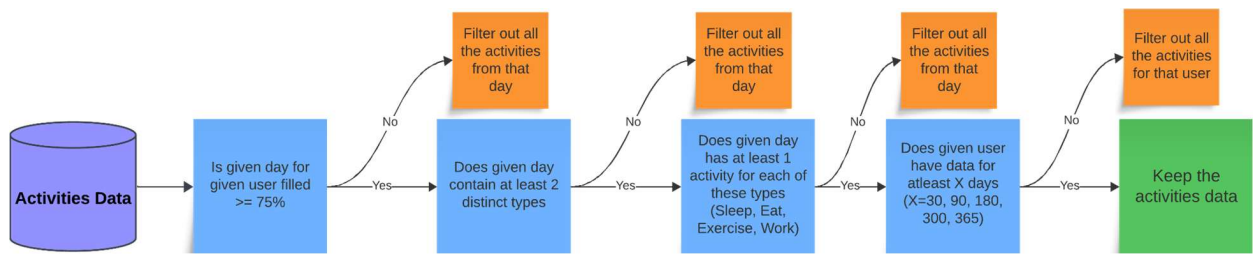

**Figure 3. Data filtering pipeline schema.**

## 3. Actigraphic inter-individual, inter-activity analysis
### a. Data Preparation

We calculated the hour times for the start and end dates of each activity. Rows with null values for essential features such as duration, endUnit, startHourTime, and midHourTime were removed. Consecutive sleep activities spanning multiple days were combined into single activities to simplify the analysis.

### b. Heatmap Construction

Median phases were used to define each individual's chronotype and to determine their amount of social jet lag. We sorted users by median values of their startHourTime and midHourTime to generate the matrixes which we then colored by cell value to make Heatmaps for Figure 1. Activities were sorted based on median onset time for selected types, and heatmaps were plotted using Seaborn's heatmap method.

## 4. Clustering Analysis
### a. Data Preparation

To analyze clustering patterns, the following metrics were calculated for each activity: duration, startHourtime, midHourtime. "Duration" captures the length of time spent on a specific activity, providing insights into a user's allocation of time across daily tasks. This is particularly relevant in understanding the internal distribution of effort and energy throughout the circadian cycle. "startHourTime" represents the hour value of the start time of an activity within a day, whereas "midHourTime" represents the hour value of the midpoint of an activity within a day. This serves as a proxy for the temporal placement of the activity relative to natural circadian phases (e.g., sleep, exercise). It helps identify whether activities align with expected chronotype patterns or deviate from typical circadian rhythms.

These features were added because they offer a quantifiable representation of circadian behaviors and provide a meaningful basis for evaluating individual chronotypes or alignment to circadian phases.

### b. K-Means Clustering

The data was grouped by user and activity type before clustering. Each group was normalized using the StandardScaler().fit_transform method. Principal Component Analysis (PCA) was applied to reduce dimensionality and to aid in visualizing the data. To determine the optimal number of clusters, the elbow method was employed by plotting the inertia values for different cluster numbers. Based on this, K-means clustering was performed using the optimal number of clusters. This process was applied to different subsets of the data, allowing for comparisons across different activity types and time periods.

### c. Hierarchical Clustering

Data was normalized using the StandardScaler().fit_transform method, similar to the K-means approach. Z-scores were computed for the normalized data, and entries with extreme Z-scores (absolute values greater than 5) were filtered out to reduce noise. Hierarchical clustering was then performed using the Ward's method for linkage, available in the scipy library, which minimizes variance within clusters. The fcluster function was used to assign cluster labels based on the dendrogram and the predetermined number of clusters.

## 5. Similarity Analysis of Clustering Techniques

To determine the similarity between clusters obtained from different clustering techniques, we performed a similarity analysis using Adjusted Rand Index (ARI) values for both the hierarchical clustering and K-means clustering results for each activity type.

## 6. ARI Values After Progressive Randomization:

We also tried to create a positive control or baseline for ARI scores. If the ARI scores after randomization still remain relatively high, it suggests that the original clustering was not much better than random. On the other hand, if the ARI after randomization becomes close to 0, it suggests that the original clustering is meaningfully better than random assignment.

Methodology:

A. Choose one clustering result: Select one clustering result from the previous analyses (we selected the clustering for Sleep activities).
B. Calculate ARI against itself: First calculate the Adjusted Rand Index (ARI) of the clustering result compared to itself. Since it's comparing a clustering to itself, the ARI should be 1 (perfect agreement).
C. Progressively randomize the clustering:
    a. Implement a loop that randomly swaps the cluster labels of two users or points.
    b. After each swap, recalculate the ARI between the original clustering and the newly randomized clustering.
D. The idea is to make the clustering increasingly more random with each iteration (cumulative randomization).
E. Repeat the process: Run this randomization loop for a specified number of iterations (we are running it for 100 times). The ARI should decrease progressively as the clustering becomes more random.

## 7. Contextual Day Analysis

The timing of key behaviors like eating and sleeping plays a crucial role in how the body's internal clock aligns with the external environment. Diving time into fixed 24-hour periods split at midnight may fail to capture the temporal dynamics of real-world behavior, where transitions between activities significantly influence circadian alignment.

We developed the concept of a contextual day, defined by the last eating activity of a day and the first sleep activity that follows it but ends in the next day. The end time of the contextual day is either the start of that sleep activity (if it begins on the next day) or 11:59:59 PM (if the sleep starts on the same day). The start time of the next contextual day begins immediately after the end of the previous contextual day, either at 12:00:00 AM or when the last contextual day ends if it's later than midnight.

## Author Contributions

SG carried out data handling and analysis, and developed figures with BS. RK and BS assisted with experimental design, analytic design. All authors contributed to the writing and editing of the manuscript.

## Conflict of Interest

BS and KS are advisors at Owaves, Inc., and RK is a founder; these three have a financial interest in Owaves, Inc. They have no other financial interests to report. SAI is an advisor at Owaves, Inc.; she has no financial interest in Owaves, Inc. SAI is a consultant for Wesper. All other authors are contracted by Owaves for their contributions to this study and have no other conflicts of interest.

# Data Availability

Owaves' data use policy does not permit us to make the data available to third parties without approval.